\documentclass{article}

\input epsf
\begin{document}

\newcommand{\lfrac}[2]{{#1}/{#2}}
\newcommand{\sfrac}[2]{{\small \hbox{${\frac {#1} {#2}}$}}}
\newcommand{\ben}{\begin{eqnarray}}
\newcommand{\een}{\end{eqnarray}}
\newcommand{\la}{\label}


\begin{center}
{\bf \Large The Gravitational Field of Massive Non-Charged Point
Source in General Relativity}

\bigskip

{\large Plamen Fiziev}
\bigskip

{\sl Department of Theoretical Physics, University of Sofia\\
Boulevard 5 James Bourchier, Sofia 1164, Bulgaria}
\medskip

E-mail: fiziev@phys.uni-sofia.bg

\end{center}

\begin{abstract}

Abstract.

Utilizing various gauges of the radial coordinate we give a
description of static spherically symmetric space-times with point
singularity at the center and vacuum outside the singularity. We
show that in general relativity (GR) there exist a two-parameters
family of such solutions to the Einstein equations which are
physically distinguishable but only some of them describe the
gravitational field of a single massive point particle with
nonzero bare mass $M_0$. In particular, the widespread Hilbert's
form of Schwarzschild solution, which depends only on the
Keplerian mass $M<M_0$, does not solve the Einstein equations with
a massive point particle's stress-energy tensor as a source. Novel
normal coordinates for the field and a new physical class of
gauges are proposed, in this way achieving a correct description
of a point mass source in GR. We also introduce a gravitational
mass defect of a point particle and determine the dependence of
the solutions on this mass defect. The result can be described as
a change of the Newton potential $\varphi_{\!{}_N}=-G_{\!{}_N}M/r$
to a modified one: $\varphi_{\!{}_G}=-G_{\!{}_N}M/
\left(r+G_{\!{}_N} M/c^2\ln{{M_0}\over M}\right)$ and a
corresponding modification of the four-interval. In addition we
give invariant characteristics of the physically and geometrically
different classes of spherically symmetric static space-times
created by one point mass. These space-times are analytic
manifolds with a definite singularity at the place of the matter
particle.

\end{abstract}

\section{Introduction}

According to the remarkable sentence by Poincar\'e, the real
problems can never be considered as solved or unsolved ones.
Instead, they are always only {\em  more or less} solved
\cite{Poincare}.

It is hard to believe that more then 88 years after the pioneer
article by Schwarzschild (1916) \cite{Schwarzschild} on the
gravitational field of a point particle with gravitational mass
$M$ there still exist serious open problems, despite of the large
number of papers and books on this problem \cite{books}.

The well known Schwarzschild \cite{Schwarzschild}  metric in
Hilbert (1917) \cite{Hilbert} gauge:

$$ds^2\!=\left(\!1-{{2M}\over\rho}\right)dt^2+{{dr^2}\over {1-{{2M}\over\rho}}}
 -\rho^2(d\theta^2+\sin^2\theta\,d\phi^2)$$
solves the {\em vacuum} Einstein equations $$G_\mu^\nu=0$$ in the
spherically symmetric static case.

It owns an {\em event horizon} at $$\rho=\rho_G=2M$$ and  a strong
hidden {\em singularity} at $$\rho=0.$$

The singularity at $\rho=0$ {\em does not describe} a massive
point particle with proper mass $M_0$, because it does not solve
the Einstein equations
\ben G^\mu_\nu = \kappa T^\mu_\nu \la{Einst}\een
in presence of matter with stress-energy tensor
$$ T^\mu_\nu \sim M_0\,\delta({\bf r}).$$
Here $\delta({\bf r})$ is the 3D Dirac function, needed to
describe the mass distribution of the point particle with proper
mass $M_0$.

According to the widespread common opinion, one is not able to use
such distributions in Einstein equations (\ref{Einst}), because
these are {\em a nonlinear} differential differential equations
\cite{YB}.

In his pioneering article Schwarzschild has used a different
radial variable $r$ and a different gauge for the spherically
symmetric static metric
$$ds^2\!=\!g_{tt}(r)^2\,dt^2\!+\!g_{rr}(r)^2\,dr^2\!-\!\rho(r)^2(d\theta^2\!+
\!\sin^2\theta\,d\phi^2).$$ Using the gauge
$$\det||g_{\mu\nu}||=1$$ he had fixed the three unknown functions
$$g_{tt}(r)=1-{{2M}\over\rho(r)}>0,\,g_{rr}(r)=-1/g_{tt}(r)<0,\,\rho(r)$$
obtaining
$$\rho(r)=\sqrt[3]{r^3+\rho_G^3}.$$

This solution has {\em no event horizon}. Its peculiar feature is
that it describes a {\em point particle with  zero radius, zero
volume, but  nonzero area} $A_\rho=4\pi\rho_G^2$.

These strange properties of the Schwarzschild solution have been
discussed by Marcel Brillouin \cite{Brillouin} in 1923 and by
Georgi Manev \cite{Manev34} in 1934.

The quantity $\rho$ has a clear geometrical and physical meaning:

$\bullet$ It is well known that $\rho$ defines the area
$$A_\rho=4\pi\rho^2$$ of a centered at $r=0$ sphere with "
area radius" $\rho$ and the length of a big circle on it
$l_\rho=2\pi\rho$.

$\bullet$ It measures, too, the curvature of the 4D-space-time and
of the 3D-space -- "curvature radius". In the spherically
symmetric case:
$${}^{{}^{(4)}}\!R={}^{{}^{(4)}}\!R(\rho),\,\,\,\,\,\,\hbox{and}\,\,\,\,\,\,
{}^{{}^{(3)}}\!R={}^{{}^{(3)}}\!R(\rho).$$ $\bullet$ One can refer
to $\rho$ as an optical "{luminosity variable}", because the
luminosity $L$ of the distant physical objects is reciprocal to
$A_\rho$: $$L\sim {1\over{\rho^2}}$$

In contrast, the {physical and geometrical meaning} of the
coordinate $r$  is not defined by the spherical symmetry of the
problem and is unknown {\em a priori}. The only clear thing is
that its value
$$r=0$$ corresponds to the center of the symmetry, where one
must place the  physical source of the gravitational field.

\section{Gauge Fixing in General Relativity}

General relativity is a gauge theory. The fixing of the gauge in
GR is described by a proper choice of the quantities
$$\bar\Gamma_\mu\!=\!-{{1}\over{\sqrt{|g|}}}g_{\mu\nu}
\partial_\lambda\left(\sqrt{|g|}g^{\lambda\nu}\right)$$
in the 4D d'Alembert operator
$$g^{\mu\nu}\nabla_\mu\nabla_\nu=g^{\mu\nu}
\left(\partial_\mu\partial_\nu-\bar\Gamma_\mu\partial_\nu\right).$$

In our problem the choice of spherical coordinates and static
metric {dictates} the form of three of them:
$$\bar\Gamma_t\!=\!0,\,\,
\bar\Gamma_\theta\!=-\!\cot\theta,\,\,\bar\Gamma_\phi\!=\!0,$$ but
the function $\rho(r)$ and, equivalently, the form of the quantity
$$\bar\Gamma_r\!=\left(\!
\ln\left({\sqrt{-g_{rr}}\over{\sqrt{g_{tt}}\,\rho^2}}\right)\!\right)^\prime$$
{are still not fixed.} Here and further on, the prime denotes
differentiation with respect to the variable $r$.

We refer to the freedom of choice of the function $\rho(r)$ as
{\em a rho-gauge freedom}  in a large sense, and to the choice of
the $\rho(r)$ function  as {\em a rho-gauge fixing}.

The strong believe in the independence of the GR results in the
choice of coordinates $x$ in the space-time ${\cal
M}^{(1,3)}\{g_{\mu\nu}(x)\}$ predisposes us to a somewhat
light-head attitude of mind towards the choice of the coordinates
for a given specific problem.

Indeed, it is obvious that physical results of any theory must not
depend on the choice of the variables and, in particular, these
results must be invariant under changes of coordinates. This
requirement is a basic principle in GR. It is fulfilled for any
{\em already fixed} mathematical problem.

Nevertheless, the change of the interpretation of the variables
may change the formulation of the mathematical problem and thus,
the physical results, because we are using the variables according
to their meaning. For example, if we are considering the
luminosity variable $\rho$ as a radial variable of the problem, it
seems natural to put the point source at the point $\rho=0$. In
general, we may obtain a physically different model, if we are
considering another variable $r$ as a radial one. In this case we
shall place the source at a different geometrical point $r=0$,
which now seems to be the natural position for the center $C$.

The relation between these two geometrical "points" and between
the corresponding physical models strongly depends on the choice
of the function $\rho(r)$, i.e. on the radial gauge. Thus,
applying the same physical requirements in different "natural"
variables, we arrive at different physical theories, because we
are solving EE under different boundary conditions, coded in
corresponding Dirac $\delta$-functions. One has to find a
theoretical or an experimental reasons to resolve this essential
ambiguity.

We shall see, that the choice of the radial coordinate in the one
particle problem in GR in the above sense is essential for the
description of its gravitational field and needs a careful
analysis. The different coordinates in a given frame are
equivalent only locally. Especially, a well known mathematical
fact is that in the vicinity of a definite singular points of
mathematical functions one must use a definite special type of
coordinates for the adequate description of the character of
singularity.

In the literature one can find a large number of useful gauges:

$\bullet$ Schwarzschild gauge (1916) \cite{Schwarzschild};

$\bullet$ Hilbert gauge (1917) \cite{Hilbert};

$\bullet$  Droste gauge (1917) \cite{Droste};

$\bullet$  Weyl gauge (1917) \cite{Weyl};

$\bullet$  Einstein-Rosen gauge (1935) \cite{Einstein}.

$\bullet$ Isotropic gauge;

$\bullet$ Harmonic gauge;

$\bullet$ Pugachev-Gun'ko-Menzel gauge (1974-76) \cite{PG}, e.t.c.

In the last case
$$\bar\Gamma_r\!=\!-{2\over r}$$  in a complete coherent way
with the flat space-time spherical coordinates. Then
$$ ds^2 \!=\! e^{2\varphi_{\!{}_N}({\bf
r})}\left(\!dt^2\!-\!{{dr^2}\over{N(r)^4}}\!\right)\!-{{r^2}\over{N(r)^2}}
\left(d\theta^2\!+\!sin^2\theta d\phi^2\right),$$ where
$N(r)\!=\!{{r}\over \rho_G}\left(1\!-\!e^{-{{\rho_G}\over
r}}\right)$.

It is remarkable that in this solution the {\em exact classical}
Newton gravitational potential
$$\varphi_{\!{}_N} ({\bf r})=-{{G_N M}\over r}$$ defines {\em an
exact GR} result.

The component $g_{tt}$ has an essentially singular point at $r=0$
 in the complex plane.

It is curious that a representation of the space-time metric of
that type was considered by Georgi Manev \cite{Manev24} already in
1924, making use of definite physical motivation.

One has to stress two apparent facts:

i) An event horizon $\rho_{{}_H}$ exists in the physical domain
{\em only} under Hilbert choice of the function $\rho(r)\equiv r$,
not in the other gauges, discussed above.

This demonstrates that the existence of black holes in the theory
strongly depends on this choice of the rho-gauge in a large sense.

ii) The choice of the function $\rho(r)$ can change drastically
the character of the singularity at the place of the point source
of the metric field in GR, because, actually, this way we are
changing the corresponding boundary conditions for Einstein
equations.

\section{Normal Coordinates for Gravitational Field of a Point
Particle in General Relativity}

Let us represent the metric $ds^2$ of the problem at hand in a
specific form:
\ben ds^2=
e^{2\varphi_1}dt^2-e^{-2\varphi_1+4\varphi_2-2\bar\varphi}dr^2-
\bar\rho^2e^{-2\varphi_1+2\varphi_2}(d\theta^2+\sin^2\theta
d\phi^2) \la{nc} \een
where $\varphi_1(r)$, $\varphi_2(r)$ and
$\bar\varphi(r)$ are unknown functions of the variable $r$ and
$\bar\rho$ is a constant -- the unit for luminosity distance
$\rho=\bar\rho\, e^{\!-\varphi_1\!+\!\varphi_2}$.

For the restriction of the gravitational action and the mechanical
action on the orbits of the group $SO(3)\bigotimes Tr_{{}_t}$ one
obtains:
\ben {\cal A}_{GR}\!=\!\!{1\over{2 G_N}}\!\!\int\!\!dt\!\!\int\!dr
\Bigl(e^{\bar\varphi}\left(\!-(\bar\rho\varphi_1^\prime)^2\!+
\!(\bar\rho\varphi_2^\prime)^2\right)
\!+\!e^{\!-\bar\varphi}e^{2\varphi_2}\Bigr),\nonumber\\ {\cal
A}_{M_0}\!=-\int\!\!dt\!\!\int\!dr\, M_0\, e^{\varphi_1}\delta(r).
\hskip 3.8truecm \la{A}\een

Thus we see that the field variables $\varphi_1(r)$,
$\varphi_2(r)$ and $\bar\varphi(r)$ play the role of {\em a normal
fields' coordinates} for our problem.

In these variables the field equations read:
\ben \bar\Delta_r \varphi_1(r)= {\frac {G_NM_0}{\bar\rho^2}}
e^{\varphi_1(r)-\bar\varphi(r)} \delta(r),\nonumber\\ \bar\Delta_r
\varphi_2(r)= {\frac {1}{\bar\rho^2}}
e^{2\left(\varphi_2(r)-\bar\varphi(r)\right)}\hskip 1.1truecm
\la{FEq}\een
where $\bar\Delta_r\!=\!e^{\!-\bar\varphi}{d\over{dr}} \left(
e^{\bar\varphi} {d \over{dr}} \right)$ is related with the radial
part of the 3D-Laplacean.

The variation of the total action with respect to the auxiliary
variable $\bar\varphi$ gives the constraint: \ben
e^{\bar\varphi}\left(-(\bar\rho\varphi_1^\prime)^2\!+\!(\bar\rho\varphi_2^\prime)^2
\right)-e^{-\bar\varphi}e^{2\varphi_2}\stackrel{w}{=}0.
\la{constraint}\een

\section{Regular Gauges and General Regular Solutions of the
Problem}

The advantage of the above normal fields' coordinates is that in
them the field equations (\ref{FEq}) are {\em linear} with respect
to the derivatives of the unknown functions $\varphi_{1,2}(r)$.
This circumstance legitimates the correct application of the
mathematical theory of distributions \cite{distrib} and makes our
normal coordinates a {\em privileged} field variables.

The choice of the function $\bar\varphi(r)$ fixes the gauge in the
normal coordinates. We have to choose this function in a way that
makes the equations (\ref{FEq}) with $\delta(r)$ function
meaningful.  We call this class {a regular gauges}. The simplest
one is the {\em basic regular gauge:}
$$\bar\varphi(r)\equiv
0\,\,\,\big(\hbox{or}\,\,\,\bar\Gamma_r=0\big).$$ Other regular
gauges describe diffeomorphysms of the fixed by the basic regular
gauge manifold ${\cal M}^{(3)}\{g_{mn}({\bf r })\}$.

Under this gauge the field equations (\ref{FEq}) acquire the
simplest ({\em linear with respect to derivatives}) form:

\ben \varphi_1^{\prime\prime}(r)= {\frac {G_NM_0}{\bar\rho^2}}\,
e^{\varphi_1(0)}\, \delta(r),\\ \varphi_2^{\prime\prime}(r)=
{\frac {1}{\bar\rho^2}}\,e^{2\varphi_2(r)}.\hskip .2truecm
\la{FEq0}\een

The constraint (\ref{constraint}) acquires the simple form, too:

\ben
-(\bar\rho\varphi_1^\prime)^2\!+\!(\bar\rho\varphi_2^\prime)^2
-e^{2\varphi_2}\stackrel{w}{=}0. \la{constraint0}\een

As one sees, the basic regular gauge $\bar\varphi(r)\equiv 0$ has
the unique property to split into {\em three independent
relations} the system of field equations (\ref{FEq}) and the
constraint (\ref{constraint}).

An unexpected feature of this {\em two parametric} variety of
solutions for the gravitational field of a point particle is that
each solution must be considered only in a definite {\em finite}
domain $r\in \big[\,0,\,r_\infty\big)$, if we wish to have a
monotonic increase of the luminosity variable in the interval
$[\rho_0,\infty)$. One easily can overcome this problem using the
regular radial gauge transformation \ben r \to r_\infty{{r/\tilde
r}\over{r/\tilde r+1 }}\la{rrgt}\een with a proper scale $\tilde
r$ of the new radial variable $r$ (Note that in the present
article we are using the same notation $r$ for different radial
variables.) We call this new regular gauge {\em a regular physical
gauge}.

The above linear fractional diffeomorfism, which carries us from
basic to the physical regular gauge, does not change the number
and the character of the singular points of the solutions in the
whole compactified complex plane of the variable $r$. The
transformation (\ref{rrgt}) simply places the point $r=r_\infty$
at infinity: $r=\infty$, at the same time preserving the initial
place of the origin $r=0$. Now the new variable $r$ varies in the
standard interval $r\in [0,\infty)$ and the regular solutions
acquire the final form. Under proper additional conditions we
obtain the solution of the problem at hand outside the source in
the form, similar to that one, written at first by Georgi Manev
\cite{Manev24}:
$$ds^2\!=\!e^{2\varphi_{\!{}_G}}
\left(dt^2\!-\!{{dr^2}\over{N_{\!{}_G}(r)^4}}\right)
\!-\!\rho(r)^2 \left(d\theta^2\!+\!\sin^2\!\theta
d\phi^2\right)\!.$$ Here we are using a modified (Newton-like)
gravitational potential: \ben \varphi_{\!{}_G}(r;M,M_0):=-{{G_N
M}\over {r+G_N M}/\ln({M_0\over M})}, \la{Gpot} \een a coefficient
$$N_{\!{}_G}(r)=\left(2\varphi_{\!{}_G}\right)^{-1}
\left(e^{2\varphi_{\!{}_G}}-1\right),$$ and an optical luminosity
variable \ben
\rho(r)\!=\!2G_NM/\left(1\!-\!e^{2\varphi_{\!{}_G}}\right)
\!=\nonumber \\{{r\!+\!G_NM/\ln({M_0\over M})}\over
{N_{\!{}_G}(r)}}. \la{rhoG} \een

As a result we reach the relations:
$$g_{tt}(0)\!=\!e^{2\varphi_1(0)}=\left(\!{{M}\over{M_0}}\!\right)^2\!<\!1.$$
The ratio $$\varrho={{M}\over{M_0}}\in (0,1)$$ describes the {\em
gravitational mass defect of the point particle}. This is the
second physical parameter in the problem at hand.

The Keplerian mass $M$ and the ratio $\varrho$ define completely
the solutions.

Outside of the source (for $r>0$) the solutions coincide with
solution in Hilbert gauge and {\em strictly respect} the Birkhoff
theorem. One must apply the Birkhoff theorem {\em only} in the
interval $\rho\in[\rho_0,\infty)$. It is remarkable that the
minimal value of the luminosity variable is \ben\rho_0=2G_N M
/(1-\varrho^2)\geq \rho_G.\la{rho_0}\een This changes the Gauss
theorem and leads to important physical consequences.

\section{Total energy of a point source and its gravitational
field}

In the problem at hand we have an extreme example of an "island
universe``. In it a privileged reference system and a well defined
global time exist. It is well known that under these conditions
the energy of the gravitational field can be defined unambiguously
\cite{books}. Moreover, we can calculate the total energy of the
aggregate of a mechanical particle and its gravitational field in
a canonical way. Indeed, the canonical procedure produces a total
Hamilton density $${\cal
H}_{tot}=\Sigma_{a=1,2;\mu=t,r}\,\pi_a^\mu\,\varphi_{a,\mu}-{\cal
L}_{tot}\!=\!{1\over{2G_N}}\left(-\bar\rho^2{\varphi_1^\prime}^2
+\bar\rho^2{\varphi_2^\prime}^2-e^{2\varphi_2}\right)+M_0
e^{\varphi_1}\delta(r).$$

For the total energy of the aggregate of the point source and its
gravitational field one obtains: \ben E_{tot}=M_0+E_{GR}=M=\varrho
M_0< M_0\,.\la{E}\een

This result follows from the relation $
E_{tot}=\int_0^{{\infty}}{\cal H}_{tot} dr\,$ and completely
agrees with {\em the strong equivalence principle} of GR.

The energy of the very gravitational field, created by the point
particle, {\em is a negative} quantity:
$$E_{GR}=E_{tot}-E_0=M-M_0<0.$$

\section{ Local Singularities of the Point Sources}

One can write down {\em an exact relativistic} Poisson equation:
\ben \Delta_r \varphi_1(r)=4\pi G_N
M_0\,\delta_g(r)=\gamma(r)\delta(r).\la{Poisson}\een This is a
specific realization of a Fock's (1964) idea  \cite{books} in our
normal fields' coordinates.

Under diffeomorphysms of the space  ${\cal M}^{(3)}\{g_{mn}({\bf r
})\}$ the singularities of the right hand side  remain unchanged.

One can distinguish the physically different solutions of Einstein
equations (\ref{Einst})  by investigation of the asymptotic  of
the coefficient
$$\gamma(r)=1/\left(4\pi\rho(r)^2\sqrt{-g_{rr}(r)}\right).$$

For our {\em regular solutions}  the limit $r\to +0$ of this
coefficient is a constant: \ben \gamma(0)={1\over {4\pi \rho_G^2}}
\left({1\over\varrho} - \varrho\right).\la{gamma}\een

\newpage
For other solutions one obtains as follows: \vskip .3truecm
\noindent Schwarzshild solution: \hskip 2.truecm
$\gamma_{{}_S}(r)\sim
{1\over {4\pi \rho_G^2}}\left({{\rho_G}\over r}\right)^{1/2}$;\\

\noindent Hilbert solution: \hskip 2.9truecm $\gamma_{{}_H}(r)\sim
{i\over {4\pi \rho_G^2}}\left({{\rho_G}\over r}\right)^{5/2}$;\\

\noindent Droste solution: \hskip 3.truecm $\gamma_{{}_D}(r)\sim
{1\over {4\pi \rho_G^2}}$;\\

\noindent Weyl solution: \hskip 3.2truecm $\gamma_{{}_W}(r)\sim
{16\over {\pi
\rho_G^2}}\left({r\over{\rho_G}}\right)^4$\,\,;\\

\noindent Einstein-Rosen solution: $\hskip 1.7truecm
\gamma_{{}_{ER}}(r)\sim {1\over {4\pi
\rho_G^2}}\left({{\rho_G}\over r}\right)^{1/2};$

\noindent Isotropic (t-r) solution: $\hskip 1.9truecm
\gamma_{{}_{ER}}(r)\sim {1\over {4\pi \rho_G^2}}\left({{x}\over
{1-x}}\right)^{1/2};$

\noindent Pugachev-Gun'ko-Menzel solution: $\hskip .1truecm
\gamma_{{}_{PGM}}(r)\!\sim\!{1\over {4\pi
\rho_G^2}}\!\left(\!{{r}\over
{\rho_G}}\!\right)^{\!2}\!\!e^{\rho_G/2r}.$

As we see:

$\bullet$ Most of the listed solutions are physically different.

$\bullet$ Only two of them: Schwarzschild and Einstein-Rosen ones,
have the same singularity at the place of the point source. As a
result, these solutions describe  diffeomorphic spaces ${\cal
M}^{(3)}\{g_{mn}({\bf r })\}$.

$\bullet$  As a result of the alteration of the physical meaning
of the variable $\rho=r$ inside the sphere of radius
$\rho_{{}_H}$, in Hilbert gauge the coefficient $\gamma(r)$ tends
to an {\em imaginary} infinity for $r\to 0$. This means that the
corresponding point source of Hilbert solution is space-like,
instead of time-like. Such property is not allowable for a source
of any physical field. This is in a sharp contrast to the {\em
real} asymptotic of all other solutions in the limit $r\to 0$.

\section{Scalar Invariants of the Riemann Tensor}

The use of scalar invariants of the Riemann tensor allows  a
manifestly coordinate independent description of the geometry of
space-time.

For our regular solutions one is able to derive the following
simple invariants:

$$I_1\!=\!{1\over{8\rho_G^2}}{{(1\!-\!\varrho^2)^4}\over{\varrho^2}}
\delta\left(\!{r\over{\rho_G}}\!\right)\!=\!{{\pi(1\!-\!\varrho^2)^3}\over{2\varrho}}
\delta_g\left(\!{r\over{\rho_G}}\!\right)\!=\hskip 1.5truecm$$ $$
-{1\over 2} R(r),$$
$$I_2\!=\!{{\Theta(r/\rho_G)\!-\!1}\over{8\rho_G^2\varrho^2}}\left(1\!-\!\varrho^2
e^{4r/\rho_G}\right)^4\!=
\!{{\rho_G^2}\over{8\varrho^2}}{{\Theta(r/\rho_G)\!-\!1}\over{\rho(r)^4}},\hskip
.77truecm$$

$$I_3\!=\!{{\Theta(r/\rho_G)}\over{4\rho_G^2}}\left(1\!-\!\varrho^2
e^{4r/\rho_G}\right)^3\!=
\!{{\rho_G}\over{4}}{{\Theta(r/\rho_G)}\over{\rho(r)^3}}. \hskip
1.9truecm$$

As seen:

$\bullet$  The invariants $I_{1,...,3}$ of the Riemann tensor are
{\em a well defined distributions.}

$\bullet$ Three of them are independent on the real axes $r\in
(-\infty,\infty)$ and this is the true number of the independent
invariants in the problem of the single point source of gravity.

$\bullet$ On the real physical interval $r\in [0,\infty)$ one has
$I_2=0$ and we remain only with two independent invariants.

$\bullet$ For $r\in (0,\infty)$ the only independent invariant is
$I_3$, as is well known from the case of Hilbert solution.

$\bullet$  The geometry of the space-time depends essentially on
both of the two parameters, $M$ and $\varrho$, which define the
regular solutions of Einstein equations.

\section{Concluding Remarks}

As we have shown, there exist infinitely many static solutions of
Einstein equations of spherical symmetry, which describe a
physically and geometrically different space-times.

Using a novel normal coordinates for gravitational field of a
single point particle with bare mechanical mass $M_0$ we are able
to describe correctly the massive static non-charged point source
of gravity in GR.

It turns out that this problem has a two-parametric family of
regular solutions, parameterized by Keplerian mass $M$ and by
gravitational mass defect ratio $\varrho={{M}\over{M_0}}\in(0,1)$.

For the regular solutions the physical values of the optical
luminosity variable $\rho$ are in the semi-constraint interval
$\rho\in [{{\rho_G}\over{1-\varrho^2}}, \infty)$.

Outside the source, i.e. for $\rho>{{\rho_G}\over{1-\varrho^2}}$,
the Birkhoff theorem is strictly respected for all regular
solutions.

For the class of our regular solutions the {\em non-physical}
interval of the optical luminosity distance
$$\rho\in[0,{{\rho_G}\over{1-\varrho^2}}],$$ which includes the
luminosity radius $\rho_{{}_H}=\rho_G$ (the event horizon), is to
be considered as {\em an optical illusion.}

For the regular solution the 3D-volume of a ball centered at the
source, is
$$Vol(r_b)={4\over 3} \pi \rho_G^3
{{12\varrho}\over{(1-\varrho^2)^4}} {{r_b}\over{\rho_G}}+ {\cal
O}_2({{r_b}\over{\rho_G}}).$$ It goes to zero {\em linearly} with
respect to the radius $r_b\to 0$.

Hence, a point source of gravity can be surrounded by a sphere
with an arbitrary small volume $Vol(r_b)$ in it and with an
arbitrary small radius $r_b$. In contrast, when $r_b\to 0$ the
area of the ball's surface has a {\em finite} limit: $${{4\pi
\rho_G^2}\over {(1-\varrho^2)^2}}> 4\pi \rho_G^2,$$ and the radius
of the big circles on this surface tends to a {\em finite} number
${{2\pi \rho_G}\over{1-\varrho^2}}>2\pi \rho_G$.

Such unusual Riemannian geometry of the point source in GR may
have an interesting physical consequences. It appeared at first in
this problem in the original Schwarzschild article
\cite{Schwarzschild} and was discussed in different aspects by
Marcel Brillouin \cite{Brillouin} and by Georgi Manev
\cite{Manev34}.

As we see, in GR we are forced to introduce an essentially {\em
new notion} for massive matter points, prescribing to them quite
unusual geometrical properties. The geometry of space-time around
such matter points must be essentially different then the geometry
around the "empty" geometrical points, or around the point with
finite density of mass-energy.

Further details and developments of present article can be found
in \cite{Fiziev}.

\end{document}